\documentclass{Interspeech2024}

\interspeechcameraready 
\usepackage{amsmath}
\usepackage{xcolor}
\usepackage{verbatim}
\usepackage{graphicx}
\usepackage{caption}
\usepackage{array}
\usepackage[space]{cite}
\usepackage{amsmath,amssymb,amsfonts}
\usepackage{graphicx}
\usepackage{textcomp}
\usepackage{subcaption}
\usepackage{xcolor}
\usepackage{tabularx}
\usepackage{textcomp}

\usepackage{multirow}
\usepackage{multicol}
\usepackage{comment}
\usepackage{amssymb}
\usepackage[vlined]{algorithm2e}
\usepackage{url}
\usepackage{float}
\usepackage{authblk}
\usepackage{etoolbox}
\usepackage{adjustbox}
\usepackage{pgf}
\usepackage{soul}
\usepackage{graphicx}
\usepackage{subcaption} 
\apptocmd{\thebibliography}{\small}{}{}
\patchcmd{\thebibliography}{\leftmargin\labelwidth}{\leftmargin\labelwidth\itemsep=0pt\parsep=0pt\topsep=0pt}{}{}

\title{PERSONA: An Application for Emotion Recognition, Gender Recognition and Age Estimation}

\name[affiliation={1}]{Devyani}{Koshal*}
\name[affiliation={1}]{Orchid}{Chetia Phukan*}
\name[affiliation={2}]{Sarthak}{Jain*}
\name[affiliation={1}]{Arun}{Balaji Buduru}
\name[affiliation={1,3}]{Rajesh}{Sharma}

\address{
  $^1$IIIT-Delhi, India,
  $^2$GGSIPU, New Delhi, India,
  $^3$University of Tartu, Estonia\\
  *equal contribution}
\email{devyani20055@iiitd.ac.in, orchidp@iiitd.ac.in, sarthakjainssjj@gmail.com}


\begin{document}

\maketitle

\begin{abstract}
Emotion Recognition (ER), Gender Recognition (GR), and Age Estimation (AE) constitute paralinguistic tasks that rely not on the spoken content but primarily on speech characteristics such as pitch and tone. While previous research has made significant strides in developing models for each task individually, there has been comparatively less emphasis on concurrently learning these tasks, despite their inherent interconnectedness. As such in this demonstration, we present \textbf{PERSONA}, an application for predicting ER, GR, and AE with a single model in the backend. One notable point is we show that representations from speaker recognition pre-trained model (PTM) is better suited for such a multi-task learning format than the state-of-the-art (SOTA) self-supervised (SSL) PTM by carrying out a comparative study. Our methodology obviates the need for deploying separate models for each task and can potentially conserve resources and time during the training and deployment phases. 
\end{abstract}

\vspace{-0.2cm}

\section{Introduction}
Speech analysis exhibits remarkable versatility across various domains, including customer service, market research, legal proceedings, healthcare, and entertainment. Within this expansive landscape, Emotion Recognition (ER), Gender Recognition (GR), and Age Estimation (AE) emerge as pivotal tasks. These tasks play an instrumental role in extracting valuable demographic insights to craft personalized recommendations and even providing the means to assess users' sentiments and satisfaction levels. \par

Previous studies have done substantial research into developing these tasks by leveraging various methodologies. Yang et al. \cite{yang2021superb} used representations from various self-supervised learning (SSL) pre-trained models (PTMs) for ER. 
Lebourdais et al. \cite{lebourdais22_interspeech} used SOTA SSL PTM WavLM representations, while Abdul et al. \cite{abdulsatar2019age} used K-NN classifier with MFCC features for GR, followed by Zazo et al. \cite{zazo2018age} using LSTM-based network for AE. \par

However, developing distinct models for each task individually presents significant resource, cost, and time implications. We tackle this challenge of resource constraints and time-intensive model development by adopting a multi-task learning strategy for the modeling and present \textbf{PERSONA} application. We explore two speech pre-trained models (PTMs) and evaluate their representations for jointly learning ER, GR, and AE. 
This approach maximises the utility of shared parameters, paving the way for more robust and scalable modeling solutions in speech processing.

\section{End-to-End Multi-Task Learning Model}
This section discusses the PTM representations used for our experiment, followed by modeling networks. Lastly, we discuss the database used and the experimental results. \par

\noindent\textbf{Pre-Trained Representations}: We use WavLM \cite{chen2022wavlm} as SSL PTM for our experiments due to its performance in SUPERB. WavLM shows SOTA performance in various speech processing tasks such as ER, ASR, deepfake detection, and so on. We use x-vector \cite{8461375}, a SOTA speaker recognition model for its topmost performance in various tasks such as ER, shout intensity prediction, and so on shown by previous works \cite{chetiaphukan23_interspeech, fukumori2023investigating}. We extract representations of 768-dimension from WavLM from the last hidden state through average-pooling and from x-vector, we extract representations of 512-dimension. 

\begin{figure*}
\centering
\includegraphics[scale=0.36]{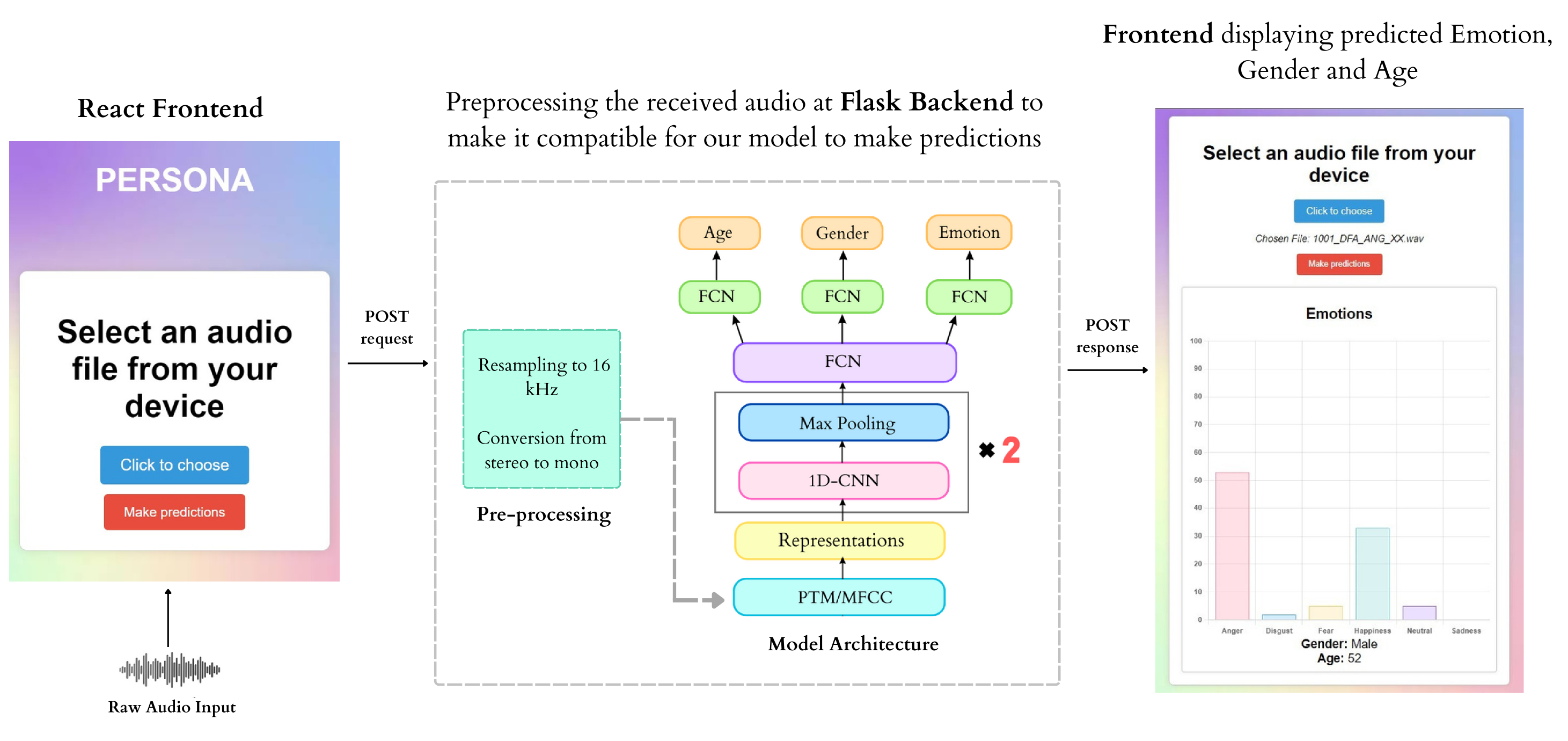}
\caption{
\textbf{PERSONA} Workflow
; Here, only the CNN model is shown}
\label{fig:obsssn}
\end{figure*} 

\noindent\textbf{Model Architecture}: We use CNN and FCN as model architectures on top of the extracted representations for our experiments. The proposed model is shown in Figure \ref{fig:obsssn}. Combination of two 1D-CNN and maxpooling layer is embedded on top of the extracted representations. This is succeeded by a fully connected network (FCN) consisting of three layers with 200, 128, and 56 neurons. The preceding layers are shared by the three tasks, followed by a task-specific head added on top, providing the task-specific output. For ER and GR, we add softmax function in the output layer that outputs the probabilities of the classes. For AE, we use a linear activation function in output layer for estimating the age. We use cross-entropy as loss function for ER, GR and mean-squared-error (MSE) for AE. For FCN, we follow the same modeling as the FCN used in the CNN model. The total loss is the summation of three losses for each specific task, and we train the model for 20 epochs with 1e-3 learning rate. We train the models in a 5-fold manner with 4-fold for training and 1-fold for testing. We leverage early stopping and learning rate decay techniques to optimise the training process.

\noindent\textbf{Database and Results}: We use CREMA-D \cite{cao2014crema} database for our experiments. We resample the audios to 16KHz before passing through the PTMs. The evaluation results of our experiments are presented in Table \ref{results}. We use accuracy as metric for ER and GR and RMSE in AE. We observed that the x-vector shows the topmost performance in the three tasks, and this behaviour can be traced to its effectiveness in capturing paralinguistic cues such as pitch, tone, etc, needed for these tasks. 


\begin{table}
\centering
\scriptsize
\setlength{\tabcolsep}{15pt}

\caption{Evaluation results; Scores given are in \%; Scores are average of 5-folds; for ER and GR, greater the better; For AE, lower the better; MFCC has been considered as baseline here} 
\label{tab:results}
\begin{tabular}{llll} \toprule
\textbf{Representation} & \textbf{Emotion} & \textbf{Gender} & \textbf{Age}\\ \midrule
\multicolumn{4}{c}{\textcolor{red}{\textbf{FCN}}} \\ \midrule
MFCC       &       47.28       &  94.02       & 7.2733   \\
WavLM       &       43.18      &    74.08  &      16.2494       \\
x-vector         &    \textbf{64.12}       &  \textbf{97.38}   &  8.1691 \\

\midrule  
\multicolumn{4}{c}{\textcolor{red}{\textbf{CNN}}} \\ \midrule

MFCC       &    43.92          & 92.55       &    8.8502  \\
WavLM       &        42.18     &   54.47   &     16.0339        \\
x-vector         &   \textbf{65.61}        & \textbf{98.52}       &  \textbf{8.6623}  \\ 
\bottomrule
\label{results}
\end{tabular}
\end{table}
\vspace{-0.4cm}

\section{PERSONA Application}
In this section, we discuss how users can interact with \textbf{PERSONA} to get the desired outputs and the complete workflow behind it. We have used React.Js for building the front end and Flask as a backend for exposing the model input and output as API. Users can upload a raw audio file in .wav or .mp3 format by clicking \textit{Click to choose} button, and then when the user clicks the button \textit{Make predictions}, the raw audio will be passed through the pre-processing stage followed by input to the model. We employ the best model CNN with x-vector representations in \textbf{PERSONA} backend. The outputs given by the model are shown at the front end that will be referred to by the user, and it is shown in Figure \ref{fig:obsssn}. \textbf{PERSONA} takes average 1sec for 1min incoming audio for inference. 


\

\vspace{-0.5cm}

\section{Conclusion} 
In this demonstration, we present, \textbf{PERSONA}, an application that predicts emotion, gender, and age from uploaded audio. Our work solves the persistent challenges associated with building individual models for ER, GR, and AE. Our work emphasizes using representations from speaker recognition PTM, x-vector instead of SSL PTM for building related multi-task applications.  

\bibliographystyle{IEEEtran}
\bibliography{main.bib}

\end{document}